\documentclass[twocolumn,tighten]{aastex701}
\submitjournal{ApJL}

\accepted{2026 May 12}

\usepackage{graphicx}
\usepackage{amsmath}
\usepackage{amssymb}

\newcommand{\HII}{\ifmmode \mathrm{H\,II}\else \ion{H}{2}\fi}

\begin{document}

\title{Compact \HII\ Regions as Clocks of Massive-Star Formation: Evidence for Long Formation Timescales}

\author[orcid=0000-0002-5055-5800,sname={Padoan}]{Paolo Padoan}
\affiliation{Department of Physics and Astronomy, Dartmouth College,Hanover, NH, USA}
\affiliation{Institut de Ci\`encies del Cosmos (ICCUB), Universitat de Barcelona (UB), Spain}
\email{ppadoan@icc.ub.edu}
\author[orcid=0000-0002-9716-1868,sname={Gieles}]{Mark Gieles}
\affiliation{Institut de Ci\`encies del Cosmos (ICCUB), Universitat de Barcelona (UB), Spain}
\affiliation{Catalan Institute of Research and Advanced Studies (ICREA), Barcelona, Spain}
\affiliation{Institut d'Estudis Espacials de Catalunya (IEEC), Barcelona, Spain}
\email{mgieles@icc.ub.edu}

\begin{abstract}

We revisit the luminosity function (LF) of compact \HII\ regions in the context of the inertial--inflow model (IIM), in which massive stars assemble over extended, mass-dependent timescales. The comparison of the compact-\HII-region LF with that of OB stars has been used to estimate the compact-\HII-phase lifetime and is often cited as evidence for the classical ``lifetime problem'' of \HII\ regions. We show that once stellar growth during the ionizing phase is included, the LF comparison instead constrains massive-star formation timescales, so the lifetime problem turns into evidence for prolonged growth. We illustrate the principle with a simple analytic model, derive revised Galactic LFs for compact \HII\ regions and OB stars from the Red MSX Source survey and the Alma Luminous Star catalogue, and fit the LFs jointly with a deterministic forward model based on stellar evolutionary tracks. The joint LF constraints imply a growth law in which the formation time is about 4 Myr for a $60\,M_\odot$ star, with an approximately square-root dependence on mass, as predicted by the IIM and supported by the numerical simulations from which it was derived. They also require the field stellar initial mass function to be a broken power law, with a slope close to Salpeter's at low masses and significantly steeper above approximately $18\,M_\odot$, as expected from the model prediction that the maximum stellar mass scales with the mass of the parent cloud. We conclude that massive stars in the Milky Way form over Myr timescales that increase with their final mass.

\end{abstract}



\section{Introduction}
\label{sec:intro}

Massive stars are rare and short-lived, yet they dominate the radiative and mechanical feedback that regulates the interstellar medium and they control the chemical evolution of galaxies. Because they contribute most of the UV/optical light of star-forming systems, interpretations of distant galaxies depend sensitively on how massive stars form and evolve. Recent JWST spectroscopy of galaxies at $z\gtrsim 11$ showing unusually high nitrogen abundances \citep[e.g.][]{Bunker+23,Castellano+24,Naidu+25} may even hint at an early role for supermassive stars \citep{Charbonnel+23,Ebihara+26}, only $\simeq 300$ Myr after the Big Bang. Despite their importance, the assembly of massive stars remains poorly understood.

\begin{figure*}
\centering
  \includegraphics[width=0.9\textwidth]{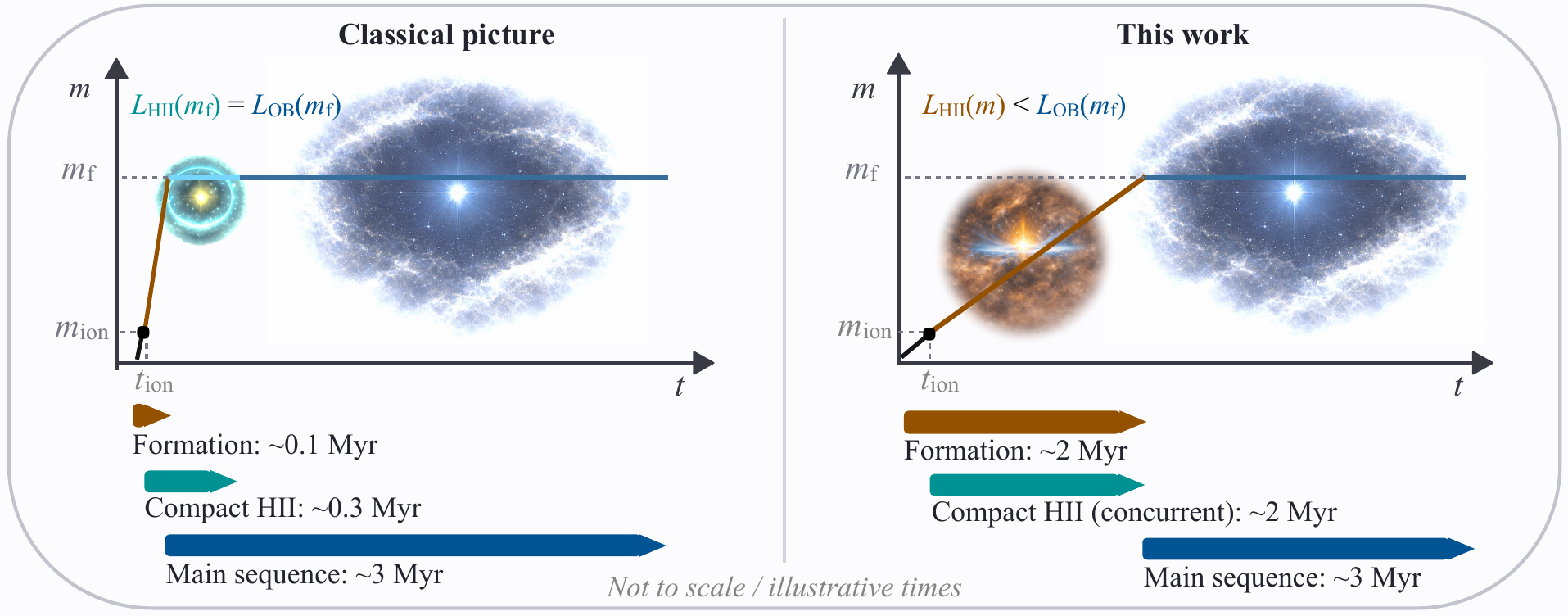}
  \caption{
  Schematic illustration of the mass growth and luminosity mapping assumed in the classical interpretation of compact \HII\ regions (left) and in the IIM picture explored in this work (right). In both panels the stellar mass $m$ increases with time until it reaches the final mass $m_{\rm f}$, after which the star evolves on the main sequence (horizontal segment). Ionising emission turns on when the growing star crosses the threshold mass $m_{\rm ion}$ at $t_{\rm ion}$ (dot). \emph{Classical picture:} the formation phase is short, and the compact-\HII\ stage is treated as a subsequent, comparatively brief phase of roughly fixed stellar mass, so that the embedded ionising source is effectively already at $m_{\rm f}$ and $L_{\HII}(m_{\rm f})=L_{\rm OB}(m_{\rm f})$. \emph{This work:} the compact-\HII\ phase occurs during continued accretion; the ionised emission is produced while $m<m_{\rm f}$, implying $L_{\HII}(m)<L_{\rm OB}(m_{\rm f})$ and motivating a comparison between the compact-\HII-region LF at luminosity $L$ and the OB-star LF at higher luminosities corresponding to the eventual $m_{\rm f}$. The coloured bars indicate approximate timescales for our reference $m_{\rm f}\simeq 60\,M_{\odot}$ case inferred in this work; they are not to scale and are intended to illustrate the ordering of phases, highlighting that in the growth scenario the compact-{\HII}-region phase is concurrent with the stellar mass assembly.
}
  \label{fig:cartoon}
\end{figure*}

In cold molecular gas the thermal Jeans mass is $\sim 1\,M_\odot$, so forming stars with $m\gtrsim 20\text{--}100\,M_\odot$ requires sustained mass transport from larger scales, increasingly shaped by feedback once the protostar becomes luminous and photoionizes its surroundings. Compact \HII\ regions (including hypercompact and ultracompact objects) mark the onset of this feedback while the system is still embedded.

Historically, compact \HII\ regions were treated mainly as a by-product of massive-star formation, and their demographics were used to infer their lifetimes. In a uniform medium an \HII\ region at $T\simeq 10^4\,$K expands at roughly the ionized sound speed and should leave the compact regime ($R\lesssim 0.1\,\mathrm{pc}$) in $\sim 10^4$ yr, yet lifetimes of $\sim 10^5$ yr are inferred from a comparison of the number of \HII\ regions to the number of OB stars, the classical ``lifetime problem'' \citep[e.g.][]{WoodChurchwell1989,WoodChurchwell1989b,Churchwell2002}. Many mechanisms were proposed to prolong the compact phase \citep[e.g.][]{Hollenbach94,Keto2002,Peters2010,GalvanMadrid2011}.

A widely used quantitative estimate was derived by \citet[][hereafter M11]{Mottram2011b}, who compared the compact-\HII-region luminosity function (LF) from the Red MSX Source (RMS) survey \citep{Urquhart2008,Mottram2011b,Lumsden2013} with the OB-star LF \citep{Reed2005}. Assuming the embedded ionizing star has already reached its main-sequence luminosity, the ratio of LFs at fixed $L$ gives the \HII-region lifetime normalized to the main-sequence lifetime,
\begin{equation}
 t_{{\HII}}(L) \;=\; t_{\rm MS}(L)\,
 \frac{\phi_{{\HII}}(L)}{\phi_{\rm OB}(L)}.
 \label{eq:t_HII_M11}
\end{equation}
M11 obtained $t_{\HII}\sim 3\times 10^5$ yr, reinforcing the lifetime problem; this value remains commonly adopted \citep[e.g.][]{Nony+24}.

The key M11 assumption---that massive stars are essentially fully formed when the embedded ionizing phase begins---matches the classical view that massive-star formation proceeds on approximately a core-collapse timescale, $\sim 10^5$ yr \citep{McKee+Tan2002,McKee+Tan2003}. In this work we adopt another view on massive star formation, leading us to a different interpretation of the nature of \HII\ regions.

In the inertial--inflow model \citep[hereafter IIM;][hereafter P20]{Padoan2020}, massive-star assembly extends to $\sim 10^6$ yr and the compact-\HII\ phase is powered by a star that is still accreting (Figure~\ref{fig:cartoon}). The appropriate LF comparison is therefore not at fixed luminosity, but between $\phi_{\HII}(L)$ and the OB-star LF at higher luminosity, $\phi_{\rm OB}(L')$ with $L'>L$, which increases the inferred $t_{\HII}(L)$ relative to equation~\eqref{eq:t_HII_M11}. In this picture the longer age is not problematic: compact \HII\ regions trace photoionized, evaporating layers of accreting structures continuously replenished by inflow rather than pressure-driven Str\"omgren spheres in a uniform medium. Here we show that the joint comparison of the compact-H\,\textsc{ii}-region and OB-star LFs constrains both the mass dependence and the overall timescale of massive-star formation in the IIM. In constructing the empirical LF and in the forward modelling, however, we broaden the compact-H\,\textsc{ii}-region sample to the full embedded population by including both compact H\,\textsc{ii} regions and massive YSOs, because our goal is to constrain the duration of the entire embedded phase of massive-star growth. This addition does not affect the mass dependence because that derivation is independent of the adopted ionization threshold (see Appendix~\ref{app:CHII_LF}).

\section{The Inertial--Inflow Framework}
\label{sec:inertial}

In this work we adopt the IIM for the origin of massive stars, introduced in P20 and recently applied to globular–cluster formation in \citet{Gieles2025}. The main features of the model needed for this work are briefly summarized in the following. 

In the IIM, massive stars are assembled by large–scale converging flows that are part of the turbulent velocity field of the star-forming cloud. These flows feed the protostar (through a circumstellar disc) from an extended, turbulent, generically unbound ``inflow region'' of parsec size. The accretion history of a star is therefore controlled primarily by the statistics of the turbulent velocity field on the scale of its mass reservoir, not by the gravitational focusing of the star itself \citep[contrary to competitive accretion;][]{Bonnell+2001a,Bonnell+2001b}.

A key result in P20 is that the average time to reach the final stellar mass, $t_{\rm form}$, increases systematically with the final stellar mass $m_{\rm f}$ as
\begin{equation}
    t_{\rm form}(m_{\rm f}) \;=\;
    \tau_0 \left( \frac{m_{\rm f}}{m_0} \right)^{\alpha},
    \label{eq:tform_param}
\end{equation}
where $m_0$ is a reference mass and $\tau_0$ is the corresponding formation time. The exponent $\alpha$ is considered to be a universal property of supersonic turbulence related to the velocity scaling, while $\tau_0$ may vary with environment. 

P20 find that the accretion rate, controlled by the inertial inflow, for an individual star is primarily stochastic, with no strong secular trend with time. This motivates a linear time–averaged growth law,
\begin{equation}
  m(t;m_{\rm f}) \simeq m_{\rm f}\, \frac{t}{t_{\rm form}(m_{\rm f})},
  \qquad 0 \le t \le t_{\rm form}(m_{\rm f}).
\label{eq:growth_law}
\end{equation}
We will refer to equations~\eqref{eq:tform_param} and \eqref{eq:growth_law} as the \emph{growth law} and to $\alpha$ and $\tau_0$ ans the \emph{growth-law parameters}.

Because the inflows that feed massive stars are generated by the turbulent cascade on scales up to the cloud (or ``outer'') scale, the IIM also predicts a natural upper limit to the stellar mass, $m_{\rm max}$, that turns out to be simply proportional the total cloud mass, $M$,
\begin{equation}
    m_{\rm max} \;\simeq\; \varepsilon_{\rm max} \, M,
    \label{eq:mmax_epsMcloud}
\end{equation}
with a dimensionless efficiency $\varepsilon_{\rm max} \simeq 2.5\times 10^{-3}$ expected to be a universal flow property in supersonic turbulence. An important consequence of equation~\eqref{eq:mmax_epsMcloud} is the steepening of the upper end of the stellar initial-mass function (IMF) of field stars, above a knee mass $m_{\rm k}=\varepsilon_{\rm max} \, M_{\rm k}$, where $M_{\rm k}$ is the minimum mass of star-forming clouds (see Appendix~\ref{app:upper_field_imf})\footnote{Related ideas based on an empirical relation between the maximum stellar mass and the stellar cluster mass have been discussed before \citep{WeidnerKroupa2006,Kroupa+13,Yan+23}, with feedback sometimes invoked as a possible physical origin. In our model, instead, the maximum stellar mass is limited by the largest turbulent inflow provided by the parent cloud, not by feedback regulation.}. We will therefore model the field IMF as a broken power law, which we will characterize with three parameters: a low-mass slope $s$ (corresponding to the slope of the IMF in individual clouds, for example the Salpeter value $s=2.35$), a high-mass slope $s_{\rm f}$ (with $s_{\rm f}>s$), and a knee mass $m_{\rm k}$. From this point on, we will sometimes use ``IMF'' as shorthand for this field IMF.

We will take equations~\eqref{eq:tform_param} and \eqref{eq:growth_law} and a broken-power-law field IMF as the working definition of the IIM adopted to compute the OB-star and compact-\HII-region LFs.

\begin{figure*}
  \centering
  \includegraphics[width=0.85\textwidth]{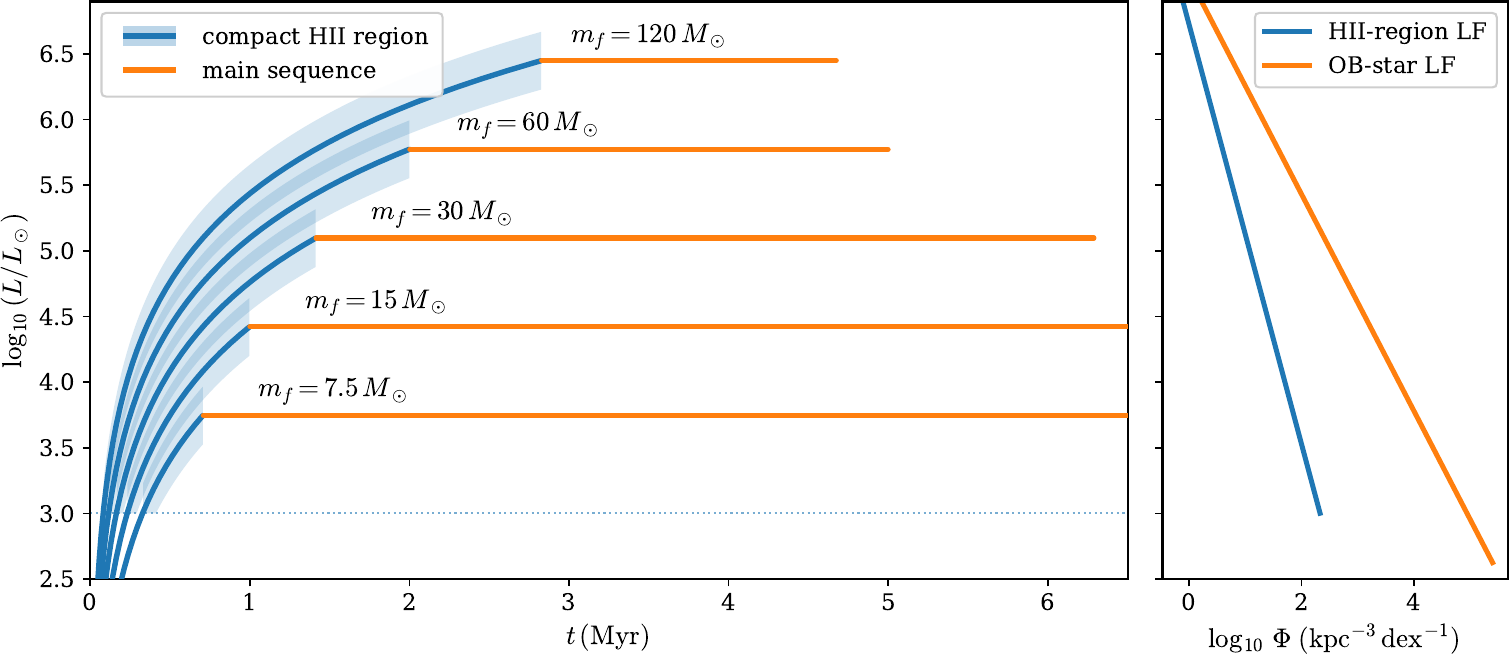}
  \caption{
Schematic mapping between stellar growth tracks in the luminosity--time plane and the observed LFs in the IIM.
\textit{Left:} example evolutionary tracks for stars of different final masses \(m_{\rm f}\). During the growth phase (blue), the luminosity increases as the star accretes; once the ionization threshold is reached (dotted horizontal line at \(\log_{10}(L/L_\odot)=3\)), the source is counted as a compact \HII\ region (blue shading) while it continues to brighten. After growth ends, the star enters the main-sequence phase (orange) at approximately fixed luminosity.
\textit{Right:} the compact-{\HII}-region and OB-star LFs (shown here schematically) can be viewed as projections of the track distribution: the {\HII}-region LF counts the time spent in each luminosity bin \emph{along the shaded portions of all tracks}, so a single luminosity bin receives contributions from a range of \(m_{\rm f}\); in contrast, the OB-star LF counts the main-sequence lifetime at the luminosity corresponding to the final mass. This figure is intended purely to illustrate the geometric origin of the LF construction in the model (not to provide a quantitative fit).
}
  \label{fig:Lt_cartoon}
\end{figure*}

\section{Predicted Luminosity Function Slopes}
\label{sec:LF_slopes}

We start with an idealised analytical derivation of the slopes of the LFs of OB stars, $\phi_{\rm OB}(L)$, and compact H\,{\sc ii} regions, $\phi_{\HII}(L)$. This derivation is intended only to illustrate how the LF shapes depend on the IMF and on the growth law. The actual constraints on those quantities are obtained later, in Section~\ref{sec:joint_model}, where the observed LFs are interpreted with the full forward model based on stellar evolutionary tracks.

To connect the IMF to $\phi_{\rm OB}(L)$ we only need to specify the relations between mass and luminosity and between mass and main-sequence lifetime. We express them as power laws with effective exponents $\gamma$ and $\delta$ such that $L \propto m^{\gamma}$ and $t_{\rm MS}\propto m^{-\delta}$ over the luminosity range of interest. To connect the IMF to $\phi_{\HII}(L)$, we further assume that compact H\,{\sc ii} regions correspond to the phase when a star is still accreting according to our growth law (see Fig.~\ref{fig:cartoon}), and has reached a mass larger than a threshold, $m_{\rm ion}$, above which it emits a significant ionizing flux. A schematic view of this construction, from the growth tracks in the \(L\)–\(t\) plane to the resulting projections into the LFs, is shown in Fig.~\ref{fig:Lt_cartoon}.

Throughout this work, the exponents \((s,s_{\rm f},\gamma,\delta,\alpha)\) are defined as positive numbers, while for the LFs we follow the usual convention $\phi_{\rm OB}(L)\propto L^{\beta_{\rm OB}}$ and $\phi_{\HII}(L)\propto L^{\beta_{\HII}}$, so that declining LFs have negative
slopes $\beta_{\rm OB}$ and $\beta_{\HII}$. We work with LFs per unit of volume and per dex in \(L\),
\(\phi(L)\equiv{\rm d}N / {\rm d}\log_{10} L\) in units of
stars kpc\(^{-3}\) dex\(^{-1}\).

\subsection{OB-star LF from the IMF}
\label{sec:OB_LF}

For clarity, we first adopt a power-law IMF for the final masses $m_{\rm f}$,
\begin{equation}
  \xi(m_{\rm f})=\frac{{\rm d}N}{{\rm d}m_{\rm f}} \;\propto\; m_{\rm f}^{-s}.
\label{eq:imf}
\end{equation}
A high-mass steepening of the IMF (Appendix~\ref{app:upper_field_imf}) may then be represented by using different effective slopes in the relevant luminosity ranges.
 
In a statistical steady state with a constant star-formation rate and a time–independent IMF, the number density of OB stars on the main sequence reflects the IMF weighted by their lifetimes. The space density per unit mass is
\begin{equation}
  n_{\rm OB}(m)
 \propto
  \xi(m_{\rm f})\, t_{\rm MS}(m_{\rm f})
  \propto m_{\rm f}^{-s-\delta}.
\end{equation}
Per dex in mass this becomes
\begin{equation}
  \frac{{\rm d}n_{\rm OB}}{{\rm d}\log m_{\rm f}}
  \propto m_{\rm f}\, n_{\rm OB}(m_{\rm f})
  \propto m_{\rm f}^{\,1-s-\delta} .
\end{equation}
Using \(m_{\rm f}\propto L^{1/\gamma}\) then gives a power–law OB LF of the
form
\begin{equation}
  \phi_{\rm OB}(L) \propto L^{\beta_{\rm OB}},
  \qquad
  \beta_{\rm OB} = \frac{1 - s - \delta}{\gamma} ,
\label{eq:beta_OB}  
\end{equation}
which can be applied separately to both the Salpeter's mass range with $s\simeq2.35$ and the steeper mass range above $m_{\rm k}$ with $s=s_{\rm f}$ (equation~\eqref{eq:sfield}).

\subsection{Compact H\,{\sc ii} region LF from the IMF}
\label{sec:CHII_LF}

To derive $\phi_{\HII}(L)$ from the single-slope IMF we adopt the growth law from equation~\eqref{eq:growth_law}, which gives a linear-in-mass track \(m(t;m_{\rm f})\) with a total duration \(t_{\rm form}(m_{\rm f})\). With this linear growth law, the accretion rate along a track of final mass $m_{\rm f}$ is constant,
\begin{equation}
\frac{dm}{dt}=\frac{m_{\rm f}}{t_{\rm form}(m_{\rm f})} .
\end{equation}
Thus, at fixed $m_{\rm f}$, the time spent in a mass interval $dm$ is proportional to $t_{\rm form}(m_{\rm f})\,dm/m_{\rm f}$. In a steady state, the compact-H\,{\sc ii}-region number density at instantaneous mass $m$ is then obtained by integrating over all stars with final mass $m_f\ge m$,
\begin{equation}
n_{\rm HII}(m)\propto \int_m^{m_{\max}} \xi(m_{\rm  f})\frac{t_{\rm form}(m_{\rm f})}{m_{\rm f
}}\,dm_{\rm f}.
\end{equation}
Using $\xi(m_{\rm f})\propto m_{\rm f}^{-s}$ and $t_{\rm form}(m_{\rm f})\propto m_{\rm f}^\alpha$ gives $n_{\rm HII}(m)\propto m^{\alpha-s}$ for $m\ll m_{\max}$, so that per dex in mass $dn_{\rm HII}/d\log m\propto m^{1+\alpha-s}$. Approximating again $L\propto m^\gamma$, we obtain
\begin{equation}
  \phi_{\HII}(L) \propto L^{\beta_{\HII}},
  \qquad
  \beta_{\HII} =
  \frac{1+\alpha - s}{\gamma} ,
\label{eq:beta_HII}
\end{equation}
(see derivation in Appendix~\ref{app:CHII_LF}).

Using the the OB–star result (equation~\ref{eq:beta_OB}), the difference btween the two slopes is 
\begin{equation}
  \Delta\beta=\beta_{\HII} - \beta_{\rm OB}
  = \frac{\alpha + \delta}{\gamma} ,
\label{eq:delta_betas}  
\end{equation}
so $\phi_{\HII}$ is always shallower (less negative $\beta$) than $\phi_{\rm OB}$, as long as $\alpha > -\delta$, that is for any growth law in which the overall formation time is not very strongly decreasing with final stellar mass ($0.5\lesssim \delta \lesssim 1.5$ in the mass range $150-10\,M_{\odot}$). Equation~\eqref{eq:delta_betas} is a central analytic result of our model: irrespective of the IMF slope ($s$ cancels out in the equation) \emph{the difference between the compact {\HII}-region and OB-star LF slopes provides an observational constraint on the exponent of the growth law of massive stars.}

\section{The Observational Luminosity Functions} \label{sec:LFs}

We derive updated LFs for OB stars and embedded massive-star sources from modern catalogues, using a common methodology that allows a consistent relative normalisation. In what follows, the OB-star LF is given in absolute Gaia magnitude $G_{\rm abs}$ (see Section~\ref{subsec:ALS_LF}), whereas the embedded-source LF is given in bolometric luminosity $L$ (Section~\ref{subsec:rms_catalogue}). In both cases, source counts are converted into space densities by assuming an exponential vertical distribution about the Galactic mid-plane with scale height $h=39$\,pc, the value inferred by M11 for compact \HII\ regions, which we adopt for both embedded sources and OB stars given their common birth environment.

To maximise the number of objects retained from the available catalogues, we include OB stars and RMS embedded sources out to heliocentric distances of 12 and 18\,kpc, respectively. We derive both LFs with the same two-step procedure: we first correct for large-scale radial variations in the observed source surface density and normalize both LFs to the source density at 3~kpc, and then apply an empirical effective-volume completeness correction (see Appendix~\ref{app:lf_corrections}).

\begin{figure*}
  \centering
  \includegraphics[width=0.49\textwidth]{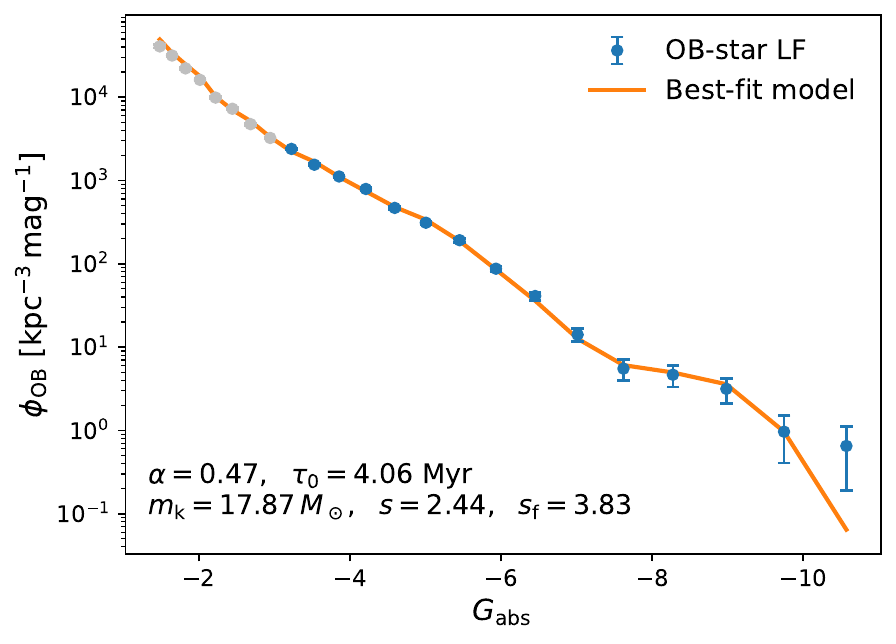}\hfill
  \includegraphics[width=0.49\textwidth]{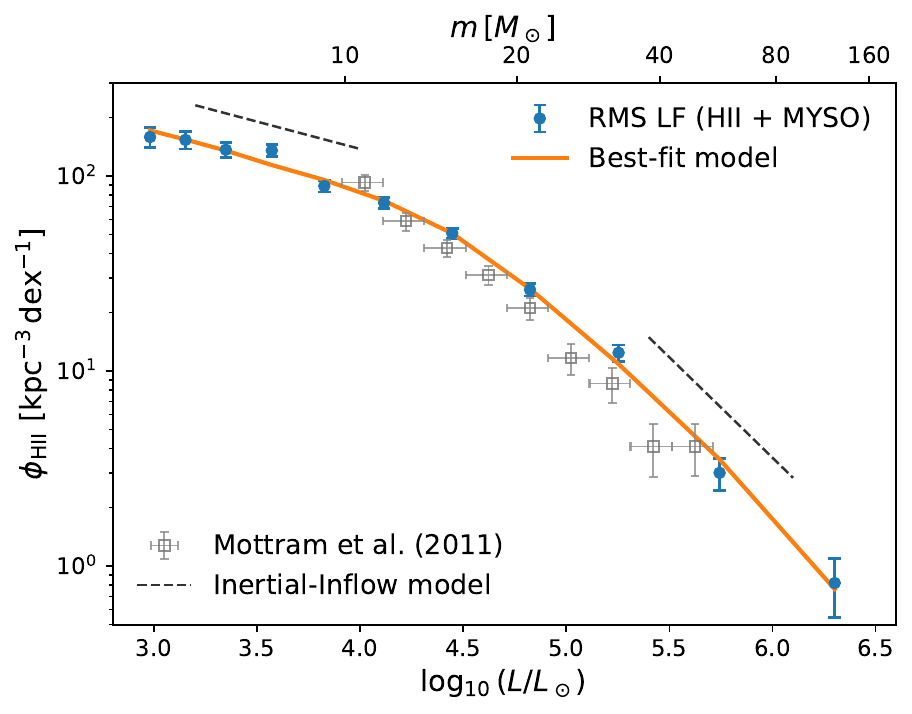}
  \caption{
\textit{Left:} Completeness-corrected OB-star LF from the ALS III catalogue (blue circles with $1\sigma$ Poisson error bars), together with the global best-fit model from the joint 5D fit described in Section~\ref{sec:joint_model} (orange curve). The horizontal axis is the absolute Gaia magnitude $G_{\rm abs}$, and the vertical axis is the number density per magnitude, $\phi_{\rm OB}$, in units of kpc$^{-3}$\,mag$^{-1}$. The text in the lower-left corner reports the best-fit parameter values, corresponding to the lowest $\chi^2$ (they differ slightly from the posterior medians quoted in Figure~\ref{fig:alpha_tau_chi2} and in Sections~\ref{subsec:growth_constraints} and \ref{subsec:joint_constraints_imf}). The light-gray points were excluded from the fit because the low-luminosity end of the LF is somewhat sensitive to the details of the completeness correction. \textit{Right:} Completeness-corrected LF of RMS embedded sources, including both compact H\,\textsc{ii} regions and massive YSOs (blue circles with $1\sigma$ Poisson error bars), together with the prediction of the same best-fit model (orange curve). Open gray squares denote the LF from M11, and the black dashed lines indicate the analytic slopes predicted by the IIM in the low- and high-mass regimes, assuming a constant star-formation rate and a stellar IMF with Salpeter's slope at low masses. The lower horizontal axis gives the bolometric luminosity, while the upper axis gives the corresponding stellar mass. The figure demonstrates that the same broken-power-law IMF, combined with the same massive-star growth law, reproduces simultaneously and accurately both the OB-star and embedded-source LFs.
}
  \label{fig:LFs}
\end{figure*}

\subsection{The OB-star LF from the ALS~III Catalog}
\label{subsec:ALS_LF}

We derive the OB-star LF using the third release of the Alma Luminous Star catalogue (ALS~III; \citealt{PantaleoniGonzalez2025}, building on \citealt{Reed2003,PantaleoniGonzalez2021}) cross-matched to \textit{Gaia}~DR3 \citep{GaiaDR3} and to the Galactic O-Star (GOS) catalogue \citep{MaizApellaniz2004}. Our sample includes all the stars from the M (likely massive stars) and I (likely intermediate-mass stars) subcatalogs, that is 15,935 stars located above the 10~kK extinction track in the color-magnitude diagram \citep[see Figure~2 in][]{PantaleoniGonzalez2025}. We take absolute $G$ magnitudes, $G_{\rm abs}$, and distances directly from the ALS~III catalog, and extinction corrections from  \textit{Gaia}~DR3, adopting $A_{\rm G}$ values from the ESP-HS pipeline (the more accurate one for hot stars) when available (10,263 stars), else from the GSP-Phot pipeline \citep{GaiaDR3GSPPhot} (2,683 stars), and assuming $A_{\rm G}=0$ when neither values exist (2,989 stars). 

Despite the obvious advantage of a more direct mass-luminosity mapping using a LF based on bolometric luminosity, $L$, we derive the LF in $G_{\rm abs}$. The bolometric correction (BC) to derive $L$ from $G_{\rm abs}$ depends on the \textit{Gaia}~DR3 effective temperature, $T_{\rm eff}$ (the spectral type is available only for a minority of the stars in the sample), which is known to be rather inaccurate \citep[see Section~3.2 in][]{PantaleoniGonzalez2025}. We found that these uncertainties propagate into an uncertainty in $L$ between 0.15~dex and 0.45~dex, also resulting in a significant uncertainty in the high-$L$ slope of $\Phi_{\rm OB}(L)$. Thus, instead of spoiling the improved photometry of the ALS~III catalog with inaccurate BCs, we have opted to derive the LF in $G_{\rm abs}$, and to apply the needed BCs to the stellar evolutionary tracks to turn the theoretical $L$ values into $G_{\rm abs}$ (see Section~\ref{sec:joint_model}). 

The resulting LF, derived with the Galactic structure and completeness corrections described in Appendix~\ref{app:lf_corrections}, is shown in the left panel of Figure~\ref{fig:LFs} (blue and gray circles with 1$\sigma$ Poisson error bars). It contains a total of 12,640 stars with $-11.0 \le G_{\rm abs} \le -1.4$ and $d\le 12$~kpc (5,958 stars in the interval $-11.0 \le G_{\rm abs} \le -3.0$ used in the next section). The left panel of Figure~\ref{fig:LFs} also shows the model LF (orange curve) derived from the five best-fit IMF and growth-law parameters (with values reported in the panel), as described in Section~\ref{sec:joint_model}.

\subsection{The Embedded-Source LF from the RMS Survey}
\label{subsec:rms_catalogue}

Our LF of embedded sources, used here as an observational counterpart of the compact-H\,{\sc ii}-region LF discussed above, is based on the Red MSX Source (RMS) survey \citep{Lumsden2013}, which provides a homogeneously selected, flux-limited sample of massive YSOs and H\,{\sc ii} regions across the Galactic plane, constructed from MSX and 2MASS colour cuts followed by extensive radio and mid-IR follow-up. Bolometric fluxes and luminosities for the RMS sources were derived by \citet{Mottram2011b} from SED fits to the near- and mid-IR data, supplemented by longer-wavelength photometry where available. 

We include both \HII\ regions and massive YSOs in the LF because our goal is to constrain the growth law of massive stars, rather than only the lifetime of compact \HII\ regions. Moreover, because the accretion rate of a massive star may fluctuate significantly with time, and because the observed appearance also depends on the line of sight, we cannot exclude that a source powered by a star with $m>m_{\rm ion}$ may at times appear observationally as a massive YSO rather than as a compact \HII\ region, for example during episodes of particularly strong accretion or along heavily obscured sightlines. In the forward model described in Section~\ref{sec:joint_model}, this combined embedded phase is used to trace the period of stellar growth.

We select all the \HII\ regions and massive YSOs with positive bolometric luminosity $L_{\rm bol}$ and heliocentric
distance $d$, which yields a working sample of
1,800 sources in the luminosity range $2.9\lesssim\log_{10}(L/L_\odot)\lesssim6.6$ and distances $d \le 18$~kpc.  Their LF, derived with the Galactic-structure and completeness corrections described in Appendix~\ref{app:lf_corrections}, is shown in the right panel of Figure~\ref{fig:LFs} (blue circles with 1$\sigma$ Poisson error bars) together with the LF from M11 (gray squares). There is good agreement in both slopes and normalizations, which is expected because we adopt the same scale height of 39\,pc as in M11, while our empirical correction is of comparable overall magnitude over the luminosity range where the two determinations overlap. The LF shows a significant steepening above a luminosity
$\log_{10}(L_{\rm k,\HII}/L_\odot) \approx 4.5$, corresponding to a `knee' mass $m_{\rm k,\HII}\approx 16\,M_{\odot}$. M11 fitted the LF with a single power law, although a change in slope is already hinted at in their data; their more limited luminosity range, however, makes the steepening much less evident. The steepening is predicted by our IIM, because the stellar IMF is expected to steepen above a certain mass, $m_{\rm k}$ (see Appendix~\ref{app:upper_field_imf}). However $m_{\rm k,\HII}$ should be smaller than $m_{\rm k}$ because in our model the star powering the H\,\textsc{ii} region has on average a mass smaller than $m_{\rm f}$ (the star is still growing).

The two dashed lines in the right panel of Fig.~\ref{fig:LFs} show the analytic predictions from Equation~(\ref{eq:beta_HII}) (see Section~\ref{sec:LF_slopes}), assuming a Salpeter slope for the stellar IMF at lower masses, $s=2.35$, a slope $s_{\rm f}=3.35$ from equation~\eqref{eq:sfield} for the steep section of the field IMF, a fiducial value of $\alpha=0.5$, and effective $L(m)$ relation exponents $\gamma=3.1$ (the average value from the MIST tracks in the mass range 6--13~$M_{\odot}$), and $\gamma=1.8$ (the average value from the MIST tracks in the mass range 40--80~$M_{\odot}$). The predicted slopes are approximately consistent with those of the observed LF in the relevant mass ranges.

\section{Joint Constraints on Massive-Star Growth and the Field IMF}
\label{sec:joint_model}

The analytic framework of Sections~\ref{sec:inertial} and \ref{sec:LF_slopes} shows that the pair of luminosity functions, $\{\phi_{\rm OB},\phi_{\rm emb}\}$, constrains the growth law of massive stars rather than a compact-H\,{\sc ii}-region lifetime at fixed luminosity. To quantify those constraints, we compare the observed OB-star and embedded-source LFs with a deterministic forward model based on the five physical parameters of the problem: the IMF knee mass $m_{\rm k}$, the low- and high-mass IMF slopes $s$ and $s_{\rm f}$, and the two growth-law parameters $\alpha$ and $\tau_0$. The details of the model construction are given in Appendix~D. Briefly, for each point in this 5D parameter space we compute both luminosity functions, compare them with the observed LFs over the adopted fitting ranges, and solve analytically for the overall normalization. The inference is therefore a genuine joint 5D scan of the full parameter set, rather than a sequence of separate fits. This formulation uses simultaneously the shapes and the relative normalization of the two LFs, which is what allows the data to constrain both the field IMF and the mass dependence of the formation time. 
The corresponding global best-fitting model is shown by the orange curves in Figure~\ref{fig:LFs}, and the resulting constraints on the growth law and the field IMF are discussed below.

\subsection{Constraints on the Growth Law}
\label{subsec:growth_constraints}

The constraints on the growth-law parameters are shown in Figure~4. Because the inference is based on the full 5D scan, the allowed region in the $(\alpha,\tau_0)$ plane is obtained after accounting for the uncertainty in the field IMF parameters, rather than by fixing the field IMF a priori. The figure shows a well-defined elongated locus, reflecting the expected covariance between the normalization and the mass dependence of the formation time: a larger value of $\alpha$ is accompanied by a larger value of $\tau_0$, and vice versa. Even so, the allowed region is compact and excludes both nearly mass-independent formation times and a very steep mass dependence.
The posterior median values shown in Figure~4 are
\begin{equation}
\alpha = 0.5^{+0.1}_{-0.1},
\qquad
\tau_0 = 4.2^{+0.9}_{-0.6}\ {\rm Myr},
\end{equation}
where the quoted uncertainties are 68\% credible intervals. Thus the data favor a growth law with a formation time of about $4$ Myr for a $60\,M_\odot$ star and an approximately square-root dependence on final mass, in agreement with the IIM and the numerical results in \citet{Padoan2020}.

\subsection{Overall Joint Constraints and the Field IMF}
\label{subsec:joint_constraints_imf}

Figure~\ref{fig:full5d} summarizes the full 5D constraints from the joint fit. The allowed regions are closed and relatively compact not only for the growth-law parameters, but also for the IMF parameters, showing that the two luminosity functions provide meaningful constraints on the entire model rather than only on a limited subset of parameter combinations. In particular, the figure shows that the high-mass IMF slope and the knee mass are both localized within restricted ranges. For the IMF parameters, the marginalized posterior distributions give
$m_{\rm k}=18\pm2\,M_\odot$, $s=2.45\pm0.09$, and
$s_{\rm f}=3.86\pm0.18$ (posterior medians with 68\%
credible intervals, approximated as symmetric). These values imply a statistically significant steepening of the field IMF above a characteristic mass scale of about $18\,M_\odot$.

\begin{figure}
   \centering \includegraphics[width=\columnwidth]{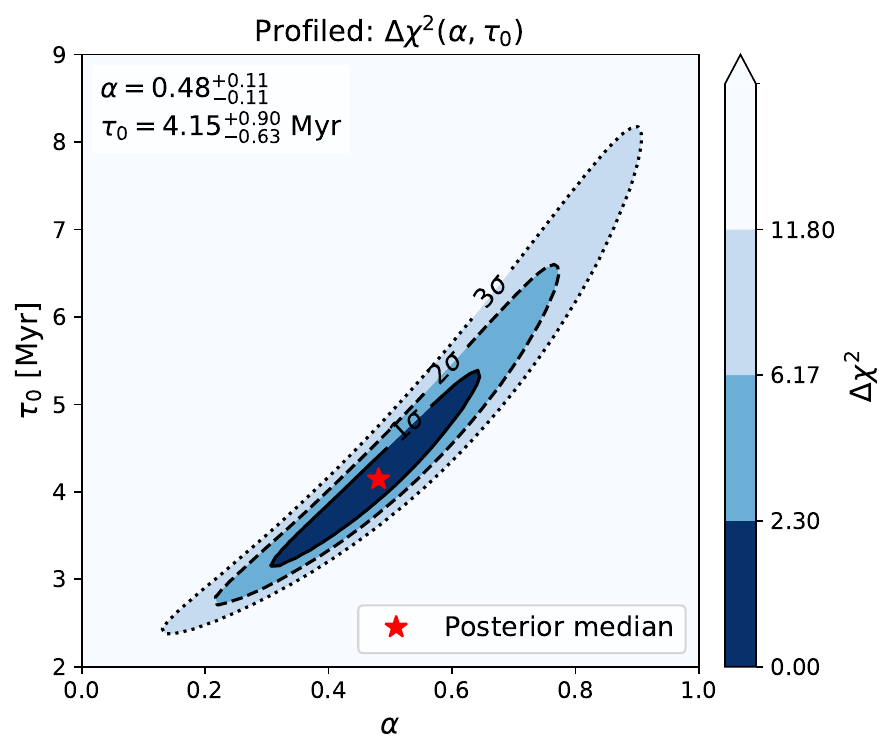}
    \caption{
    Constraint on the massive-star growth-law parameters $\alpha$ and $\tau_{0}$ (see equation~\eqref{eq:tform_param}) from the full 5D joint fit of the OB-star and embedded-source LFs. The colored map shows the profiled $\Delta\chi^{2}(\alpha,\tau_{0})$, obtained after minimizing over the three IMF parameters at each point in the $(\alpha,\tau_{0})$ plane. Contours correspond to the joint 1, 2, and $3\sigma$ confidence regions for two parameters, with $\Delta\chi^{2}=2.30$, 6.17, and 11.8. The red star marks the posterior median. The values shown in the upper-left corner give the corresponding 68\% credible intervals, $\alpha=0.48^{+0.11}_{-0.11}$ and $\tau_{0}=4.15^{+0.90}_{-0.63}\,\mathrm{Myr}$.
    }
    \label{fig:alpha_tau_chi2}
\end{figure}

\begin{figure}
  \centering
  \includegraphics[width=\columnwidth]{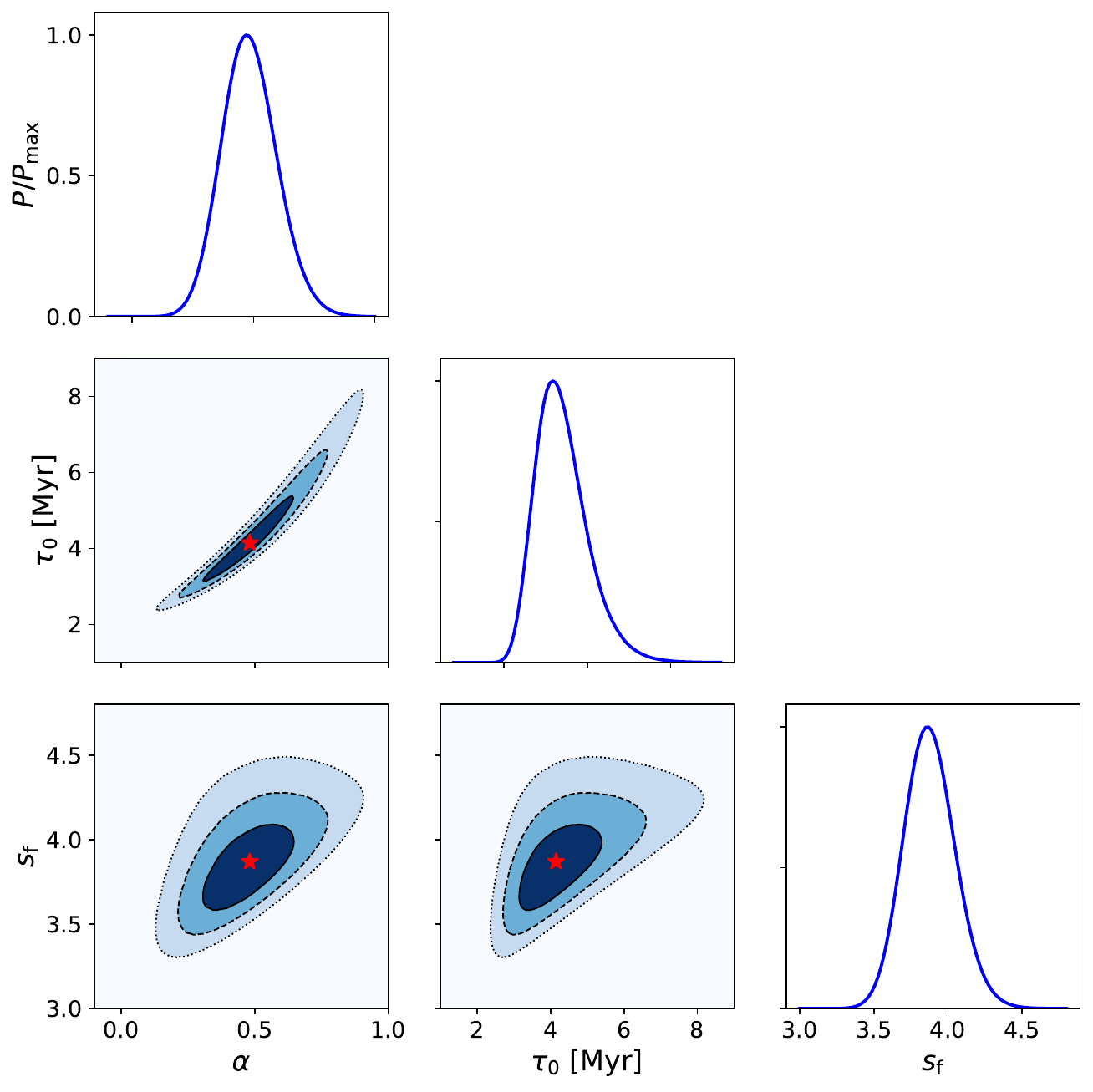}
  \includegraphics[width=\columnwidth]{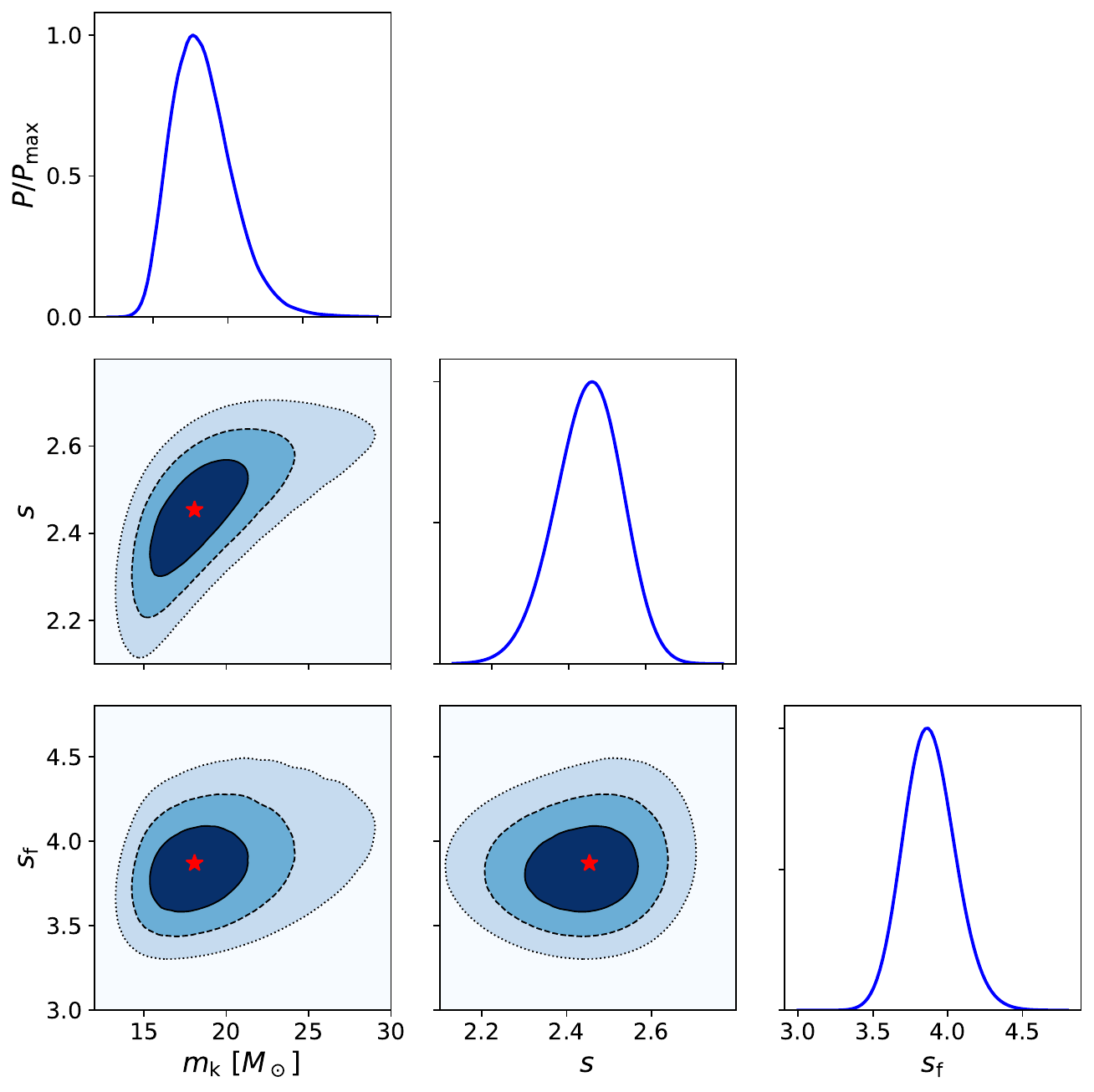}
  \caption{
    Summary of the full 5D constraints from the joint fit of the OB-star and embedded-source LFs. The diagonal panels show the marginalized one-dimensional posterior distributions of the five model parameters, normalized to their peak values. The off-diagonal panels show the corresponding two-dimensional profiled confidence regions, with contours at $\Delta\chi^2=2.30$, 6.17, and 11.8, corresponding to the 1, 2, and 3$\sigma$ confidence levels for two parameters. Red stars mark the posterior medians. The closed, relatively compact contours indicate that the joint fit provides meaningful constraints on all five parameters.
  }
  \label{fig:full5d}
\end{figure}

The evidence for a broken power-law IMF is decisive in the joint analysis. Relative to the best single-power-law model from the corresponding 3D scan, the best broken-power-law model from the 5D scan lowers the minimum $\chi^2$ from 111.7 to 34.3, a decrease of $\Delta\chi^2=77.4$ despite the two additional IMF parameters. The same conclusion is reached with standard model-selection criteria that penalize the larger parameter space, with $\Delta{\rm AIC}=73.4$ and $\Delta{\rm BIC}=70.9$ in favor of the broken-power-law model. These differences correspond to relative support for the single-power-law model at the level of only $\sim 10^{-16}$, so the steepening of the IMF is a robust requirement of the data.

The strong and localized steepening inferred here, with a knee near $18\,M_\odot$ and a high-mass slope significantly steeper than Salpeter's, is, to our knowledge, a new observational result. It is also qualitatively consistent with a fundamental prediction of the IIM: the approximate proportionality between maximum stellar mass and parent-cloud mass implies that the field IMF must steepen above a characteristic mass scale (Appendix~\ref{app:upper_field_imf}). Alternative ideas based on cluster-scale sampling effects appear to predict a weaker and more gradual steepening than the one inferred here \cite[e.g.][]{WeidnerKroupa2006}, although we do not claim that our physical interpretation of the break is necessarily unique.

\section{Conclusions}
\label{sec:conclusions}

We have revisited the comparison between the LFs of compact \HII\ regions and OB stars in the framework of the IIM, in which massive stars assemble over extended, mass-dependent formation timescales. Using revised observational LFs and a deterministic joint forward model based on a broken-power-law field IMF, a growth law, and stellar evolutionary tracks, we have obtained strong constraints on the growth law of massive stars and on the shape of the field IMF. We stress that while the exponent $\alpha$ of the growth law is expected to reflect a universal property of supersonic turbulence, the timescale $\tau_0$ may vary with environment and in particular may be smaller in regions of higher gas surface density \citep[see the scaling  derived in][]{Gieles2025}, but still of order Myr. Our main results are as follows:

\begin{enumerate}
\item We have derived updated LFs for embedded sources (compact \HII\ regions and massive YSOs; 1,800 sources in total) and OB stars (12,640 stars in total, of which 5,958 lie in the fitting range used to constrain the model LF) from the RMS and ALS\,III catalogues, respectively, correcting both for large-scale radial structure and incompleteness, and placing them on a common normalization at 3~kpc.

\item A deterministic forward model reproduces simultaneously the observed OB-star and embedded-source LFs over the fitted luminosity ranges, including both their shapes and their relative normalization. This joint agreement is non-trivial because it requires the same IMF, growth law, and birthrate normalization to account for both optically-revealed stars and embedded-source populations.

\item Jointly modelling the OB-star and embedded-source LFs constrains the massive-star growth law,
$t_{\rm form}(m_{\rm f})=\tau_0(m_{\rm f}/m_0)^{\alpha}$. Marginalizing over the IMF parameters, we obtain $\alpha = 0.5\pm0.1$ and $\tau_0 = 4.2^{+0.9}_{-0.6}\,{\rm Myr}$ for $m_0=60\,M_\odot$, consistent with the IIM expectation of an approximately square-root dependence of formation time on final mass.

\item The IMF is also constrained by the joint fit, with
$m_{\rm k}=18\pm2\,M_\odot$, $s=2.45\pm0.09$, and $s_{\rm f}=3.86\pm0.18$. The statistical preference for a broken power-law IMF over a single power law is overwhelming.

\item The strong and localized steepening of the field IMF above a characteristic mass is a distinctive prediction of the IIM and, to our knowledge, a new observational result.

\item In this framework, the long ``lifetimes'' of embedded sources inferred from LF comparisons are naturally reinterpreted as evidence that massive YSOs and compact \HII\ regions trace an extended phase of ongoing mass assembly in which accretion and photoionization coexist, rather than an expansion or confinement timescale problem.
\end{enumerate}

Our scenario is broadly compatible with earlier ideas in which compact H\,{\sc ii} regions are associated with ongoing accretion or with photoionized dense gas near the star \citep[e.g.][]{Keto2002,Peters2010}, although those models were not developed in the context of the long growth times predicted by the IIM and constrained by the LFs. These long growth times may also provide a natural solution to the zero-age main-sequence discrepancy in massive stars \citep[][]{Bernasconi+Maeder96,Holgado+20}, a connection that we will explore in forthcoming work.

The simulations on which the IIM is based do not yet include ionizing feedback; they describe only the mass assembly driven by turbulent inflow. When ionizing radiation is added to this framework, the relevant structure is unlikely to resemble a pressure-confined Str\"omgren sphere. A more natural outcome is that the ionized gas traces photoevaporating surfaces of the dense accreting structures, such as a disk or pseudodisk, and nearby filaments. The resulting morphology may be complex and time-dependent, but the dominant free-free emission should arise from the highest-emission-measure gas close to the star, so the main observed \HII\ region can remain compact even if ionizing photons leak through lower-density channels. Previous studies of massive-star formation have shown that radiative feedback and accretion can coexist in sufficiently anisotropic flows, with radiation and ionized gas escaping through low-density cavities while accretion continues through shielded structures \citep[e.g.][]{Krumholz2009,Cunningham2011,Rosen2016,KuiperHosokawa2018,Tanaka2017}. Those calculations, however, generally address more idealized core-accretion geometries and much higher accretion rates than the Myr-scale growth inferred here. Testing the IIM scenario therefore requires a new class of radiation-hydrodynamic simulations in which ionizing feedback is coupled to turbulent, large-scale inflows. Such calculations are now being developed and will be presented in future work, but we do not expect stellar feedback to alter the main qualitative conclusion that massive stars form over Myr timescales that increase with their final mass.


\begin{acknowledgments}
We thank the referee for constructive comments that helped improve the clarity and presentation of the paper. PP acknowledges support by the US National Science Foundation under Grant AST 2408023. MG acknowledge financial support from the grants PID2024-155720NB-I00, CEX2024-001451-M funded by MCIN/AEI/10.13039/501100011033 (State Agency for Research of the Spanish Ministry of Science and Innovation). This research was supported by the International Space Science Institute (ISSI) in Bern, through ISSI International Team project 'The Origin of Multiple Populations in Globular Clusters’ (ISSI Team project \#25-636).
\end{acknowledgments}

\appendix

\section{The upper end of the field IMF}
\label{app:upper_field_imf}

Because $m_{\rm max}$ is proportional to the total mass of the star–forming cloud (equation~\ref{eq:mmax_epsMcloud}), the \emph{field} IMF, which is the outcome of star-formation from many clouds with different mass $M$, cannot be a single power law: it must steepen at masses where the finite $m_{\rm max}$ of typical clouds becomes important. To model this, we introduce a \emph{knee mass} $m_{\rm k}$, defined as the typical maximum stellar mass in the lowest--mass clouds capable of forming stars, corresponds to a characteristic cloud mass
\begin{equation}
  M_{\rm k} \;\equiv\; \frac{m_{\rm k}}{\varepsilon_{\rm max}}.
\end{equation}
For the purpose of this derivation we simply assume that clouds with mass $<M_{\rm k}$ do not contribute to star formation due to their large virial parameters \citep[e.g][]{Heyer+2001,Evans+21} and thus negligible star-formation rate \citep{Padoan+Nordlund11,Padoan+12,Padoan+17}. Above $M_{\rm k}$, we assume that star-forming clouds follow a power--law mass function,
\begin{equation}
  \frac{{\rm d}N_{\rm cloud}}{{\rm d}M} \;=\;
  C_{\rm cloud}\,M^{-\beta},
\end{equation}
as well documented for Galactic molecular clouds  \citep[e.g.][]{Heyer+2001,Roman-Duval+2010,Miville-Deschenes+2017,Colombo+2019} and for molecular clouds in nearby galaxies \citep[e.g.][]{Rosolowsky2005,Gratier+2012,Colombo+2014,Utomo+2015,Faesi+2018,Rosolowsky+2021}. The characteristic value of the exponent is $\beta \simeq 2$, but with significant variations in different environments (often larger at larger galactocentric distances or in inter-arm regions).

To evaluate the steepening, we assume that within a given cloud the stellar IMF (the distribution of final stellar masses, $m_{\rm f}$) at high mass is a power law,
\begin{equation}
  \xi(m_{\rm f})=\frac{{\rm d}N}{{\rm d}m_{\rm f}} \;\propto\; m_{\rm f}^{-s},
\label{eq:imf_app}
\end{equation}
truncated at the maximum mass $m_{\max}(M)=\varepsilon_{\rm max}M$. The \emph{field} IMF at mass $m_{\rm f}$ is then obtained by summing over all clouds that can host such stars, namely all clouds with
$M \ge M_{\rm k}$:
\begin{equation}
  \xi_{\rm field}(m_{\rm f}) \;\propto\;
  \xi(m_{\rm f})
  \int_{{M_{\rm k}}}
  M^{-\beta}\,{\rm d}M.
\end{equation}
Thus, for $m_{\rm f} \gtrsim m_{\rm k}$ we obtain
\begin{equation}
  \xi_{\rm field}(m_{\rm f})
  \;\propto\;
  m_{\rm f}^{-s}\,m_{\rm f}^{1-\beta}
  \;=\;
  m_{\rm f}^{-(s + \beta-1)},
\end{equation}
so the effective high--mass slope of the field IMF is
\begin{equation}
  s_{\rm f} \;=\; s + \beta - 1\simeq\; 3.35,
  \label{eq:sfield}
\end{equation}
where the approximate value assumes $\beta\simeq 2$ and $s\simeq 2.35$ \citep{Salpeter1955}.\footnote{The steepening of the field IMF due to a dependence of the maximum stellar mass on the cluster mass was discussed in \citet{WeidnerKroupa2006}. However, the steepening is less pronounced and model dependent, with the largest value of $s_{\rm f}=3.0$ starting from $\beta=2.35$ (where $\beta$ is the slope of the cluster mass function), and the steepening applies to all stars above 1~$M_{\odot}$.}

\section{The slope of the H\,{\sc ii}-region LF}
\label{app:CHII_LF}

As a star grows, it eventually becomes massive enough to ionize its surroundings. We define an ionization threshold
\(m_{\rm ion}\), such that a star with instantaneous mass
\(m(t;m_{\rm f}) \ge m_{\rm ion}\) drives a compact H\,{\sc ii} region. With the linear growth law (equation~\eqref{eq:growth_law}), the ionizing phase for a star of final
mass \(m_{\rm f} > m_{\rm ion}\) begins when \(m(t;m_{\rm f}) =
m_{\rm ion}\) and ends when \(m(t;m_{\rm f}) = m_{\rm f}\). The
corresponding start time is
\begin{equation}
  t_{\rm ion}(m_{\rm f}) = t_{\rm form}(m_{\rm f}) \,
                           \frac{m_{\rm ion}}{m_{\rm f}},
\end{equation}
so the duration of the compact H\,{\sc ii} phase is
\begin{equation}
  t_{{\HII}}(m_{\rm f}) =
  t_{\rm form}(m_{\rm f}) - t_{\rm ion}(m_{\rm f})
  = t_{\rm form}(m_{\rm f}) \left[ 1 - \frac{m_{\rm ion}}{m_{\rm f}} \right].
\label{eq:t_CHII}  
\end{equation}

The growth law (equation~\eqref{eq:growth_law}) implies a constant accretion rate along each track,
\begin{equation}
  \frac{{\rm d}m}{{\rm d}t}
  = \frac{m_{\rm f}}{t_{\rm form}(m_{\rm f})} .
\end{equation}
During the compact H\,{\sc ii} phase the instantaneous mass runs from
$m=m_{\rm ion}$ to $m=m_{\rm f}$, so at fixed $m_{\rm f}$ the time
spent in a mass interval ${\rm d}m$ is
${\rm d}t = t_{\rm form}(m_{\rm f})\,{\rm d}m/m_{\rm f}$.
The total compact H\,{\sc ii} lifetime for that star is
$t_{{\HII}}(m_{\rm f})$, from equation~\eqref{eq:t_CHII}, so the fraction of the compact
H\,{\sc ii} lifetime spent in ${\rm d}m$ is
${\rm d}t / t_{{\HII}}(m_{\rm f})$. Therefore, in a steady state with a constant star-formation rate, the number density of compact H\,{\sc ii} regions contributed by stars of final mass $m_{\rm f}$ in the interval $[m,m+{\rm d}m]$ is 
\begin{equation}
  {\rm d}n_{\HII}(m\mid m_{\rm f})
  \propto \xi(m_{\rm f})\,t_{{\HII}}(m_{\rm f})\,
           \frac{{\rm d}t}{t_{{\HII}}(m_{\rm f})}
  \propto \xi(m_{\rm f})\,
           \frac{t_{\rm form}(m_{\rm f})}{m_{\rm f}}\,{\rm d}m .
\end{equation}
The factor $t_{\HII}(m_{\rm f})$
cancels out, so the result is independent of the choice of $m_{\rm ion}$
apart from the requirement that $m\ge m_{\rm ion}$ (this also implies that the predicted slope is independent of whether we include the whole growth phase or only its ionizing portion).

Integrating over all final masses that can reach a given instantaneous
mass $m$ then gives the steady--state number density per unit mass,
\begin{equation}
  n_{\HII}(m) \propto
  \int_{m}^{m_{\max}}
  m_{\rm f}^{-s}\,
  \frac{t_{\rm form}(m_{\rm f})}{m_{\rm f}}\,
  {\rm d}m_{\rm f} ,
  \qquad (m \ge m_{\rm ion}) ,
\end{equation}
where $m_{\max}$ is the upper cutoff of the IMF in a given cloud. Substituting
$t_{\rm form}(m_{\rm f})\propto m_{\rm f}^{\alpha}$ yields an
integrand $\propto m_{\rm f}^{\alpha-s-1}$, so that
\begin{equation}
  n_{\HII}(m) \propto
  \frac{
    m_{\max}^{\,\alpha-s}
    - m^{\alpha-s}
  }{\alpha - s},
  \qquad (\alpha \neq s,\; m \ge m_{\rm ion}) .
\end{equation}
This expression explicitly retains both integration limits and shows
that, in general, the compact H\,{\sc ii} LF is an integral transform
of the IMF and the growth law, rather than a pure power law.

For realistic parameters \(\alpha < s\) the exponent \(\alpha - s\) is
negative. For masses well inside the allowed range,
\(m_{\rm ion} \lesssim m \ll m_{\max}\), the second term in the
numerator dominates in absolute value and the scaling reduces to
\begin{equation}
  n_{\HII}(m) \propto m^{\alpha-s} .
\end{equation}
Per dex in instantaneous mass this becomes
\begin{equation}
  \frac{{\rm d}n_{\HII}}{{\rm d}\log m}
  \propto m\, n_{\HII}(m)
  \propto m^{\,1+\alpha-s} .
\end{equation}
Approximating the compact H\,{\sc ii} region luminosity as that of the ionizing star, \(L\propto m^{\gamma}\), we obtain
\begin{equation}
  \phi_{\HII}(L) \propto L^{\beta_{\HII}},
  \qquad
  \beta_{\HII} =
  \frac{1+\alpha - s}{\gamma} ,
\label{app:beta_HII}
\end{equation}
which can be applied separately to both the shallower range with $s=2.35$ and the steeper range with $s=s_{\rm f}$ (equation~\ref{eq:sfield}).

\section{Luminosity-Function Methodology}
\label{app:lf_corrections}

The OB-star and embedded-source LFs are derived with the same two-step procedure. First, we correct for large-scale radial variations in the observed source surface density across the Galactic star-forming disk, so that these variations do not bias the inferred LF slope. This step also places both LFs on a common absolute normalization, defined as the source density at a heliocentric distance of 3~kpc. Second, we correct for incompleteness using an empirical effective-volume method derived directly from the data. The resulting LFs are therefore corrected for both large-scale radial structure and the finite depth of the catalogues: per unit magnitude in $G_{\rm abs}$ for OB stars, and per unit dex in luminosity for embedded sources.

\subsection{Galactic-Structure Correction and Normalization to 3~kpc}
\label{app:galactic_structure_correction}

For each catalogue, we derive an empirical radial correction factor, $g(R)$, where $R$ is the heliocentric distance projected onto the Galactic plane. The function $g(R)$ is estimated from source counts in a restricted magnitude or luminosity range chosen to be well populated and only weakly affected by incompleteness, and is normalized so that
\begin{equation}
g(3\,\mathrm{kpc}) = 1 \, .
\end{equation}
When required by the data, the correction is estimated in more than one reference range and combined into a single smooth function of $R$.

The function $g(R)$ should be understood as an empirical correction for the radial variation of the \emph{observed} source density, irrespective of its physical origin. For the embedded-source sample, this variation is expected to trace primarily the large-scale structure of the Galactic star-forming disk. For the OB-star sample, however, the decline in counts beyond a few kiloparsecs is likely enhanced by extinction along spiral-arm sightlines, in addition to any true variation in the underlying stellar density. For the purpose of LF construction, this distinction is not crucial: in both cases, $g(R)$ removes a large-scale distance-dependent bias before the incompleteness correction is estimated.

For each cumulative distance limit $R$, we first compute the LF in the usual way from all sources with heliocentric distance $d \le R$, assuming an exponential vertical distribution with scale height $h$, so that the effective survey volume is
\begin{equation}
V(R) = 2h\,\pi R^2 \, .
\end{equation}
We then correct the normalization of that distance-limited LF by dividing by $g(R)$. Thus, for the embedded-source LF,
\begin{equation}
\phi_g(L;R)=\frac{\phi(L;R)}{g(R)} \, ,
\end{equation}
where $\phi(L;R)$ is expressed per unit dex in luminosity, while for the OB-star LF,
\begin{equation}
\phi_g(G_{\rm abs};R)=\frac{\phi(G_{\rm abs};R)}{g(R)} \, ,
\end{equation}
where $\phi(G_{\rm abs};R)$ is expressed per unit magnitude in $G_{\rm abs}$.

These corrected distance-limited LFs are then used in the completeness correction described in Appendix~\ref{app:effective_volume_completeness}. The maximum heliocentric distances adopted for the catalogues are 12~kpc for the OB-star sample and 18~kpc for the embedded-source sample.

\subsection{Empirical Effective-Volume Completeness Correction}
\label{app:effective_volume_completeness}

After the Galactic-structure correction, we estimate the completeness correction empirically from the same catalogue. At fixed luminosity or absolute magnitude, the inferred number density should be approximately independent of the adopted maximum heliocentric distance as long as the sample remains effectively complete at that value. Once the distance limit exceeds the completeness horizon, the inferred density decreases because the sampled volume grows faster than the number of detected sources.

For a sequence of distance limits $R$, we therefore compute the structure-corrected, distance-limited LF, $\phi_g(L;R)$, and define an effective maximum distance $R_{\rm eff}(L)$ by the condition
\begin{equation}
\phi_g[L;R_{\rm eff}(L)] = \max_R \, \phi_g(L;R) .
\end{equation}
Equivalently, one may define an effective volume
\begin{equation}
V_{\rm eff}(L) = V[R_{\rm eff}(L)] .
\end{equation}
This is the largest effective volume over which the catalogue remains consistent with being complete at luminosity $L$.

In principle, the LF could be evaluated directly from $\phi_g[L;R_{\rm eff}(L)]$. In practice, because $R_{\rm eff}$ becomes small at the faint end, that estimator would use only a limited fraction of the data and would therefore be unnecessarily noisy. Instead, we use the effective-volume construction to derive a smooth multiplicative completeness correction relative to a single large reference sample. Specifically, for each catalogue we first compute a raw structure-corrected LF,
\begin{equation}
\phi_{g,0}(L) \equiv \phi_g(L;R_0) ,
\end{equation}
where $R_0$ is the maximum distance adopted for the catalogue. We then define the completeness factor in coarser luminosity bins as
\begin{equation}
C(L_k) \equiv \frac{\max_R \, \phi_g(L_k;R)}{\phi_{g,0}(L_k)} \ge 1 .
\end{equation}
Treating completeness as a smooth function of luminosity, we interpolate these discrete values to obtain a continuous correction $C(L)$, and define the final LF as
\begin{equation}
\phi(L) = C(L)\,\phi_{g,0}(L) .
\end{equation}
The same factor is applied to the Poisson uncertainties of the raw LF.

The same procedure is applied to both catalogues, using $G_{\rm abs}$ for the OB-star LF and $\log_{10}(L/L_\odot)$ for the embedded-source LF. Because the Galactic-structure correction is applied first, the completeness correction acts only on the residual loss of sources with distance and does not absorb the large-scale radial variation of the observed source density.

The LFs shown in Figure~\ref{fig:LFs} are the result of this common two-step procedure and can therefore be compared directly in the joint forward modelling described in Appendix~\ref{app:joint_model}.

\section{Deterministic forward model for the luminosity functions}\label{app:joint_model}

The forward model includes the same physical ingredients introduced in Sections~\ref{sec:inertial}--\ref{sec:LF_slopes}, but uses a more detailed treatment of stellar evolution and a deterministic construction of the luminosity functions. Its ingredients are:
\begin{itemize}
  \item a broken-power-law field IMF with knee mass $m_{\rm k}$ and slopes $s$ and $s_{\rm f}$ below and above the knee;
  \item a formation-time law $t_{\rm form}(m_{\rm f})$ of the form of equation~\eqref{eq:tform_param}, where $\alpha$ and $\tau_0$ are the growth-law parameters to be constrained;
  \item a growth track $m(t;m_{\rm f})$ that increases linearly in mass from a seed mass $m_{\rm seed}=1\,M_{\odot}$ to the final mass $m_{\rm f}$ over $t_{\rm form}(m_{\rm f})$, as in equation~\eqref{eq:growth_law};
  \item a definition of the embedded phase (massive YSOs and \HII\ regions) as the whole growth period between $m=m_{\rm seed}$ and $m=m_{\rm f}$;
  \item an $L(t;m_{\rm f})$ relation based on the \emph{Modules for Experiments in Stellar Astrophysics} (MESA) Isochrones and Stellar Tracks (MIST; \citealt{Dotter2016,Choi2016}), including post-main-sequence phases and the appropriate bolometric corrections to compute $G_{\rm abs}$ for the OB-star LF model;
  \item an initial luminosity of the optically revealed star at the end of the formation period, $L(t=t_{\rm form};m_{\rm f})$, taken from the evolutionary track at an age shifted relative to the zero-age main sequence to account for the fraction of core hydrogen already burned during the growth phase;
  \item the observational LFs from Section~\ref{sec:LFs}.
\end{itemize}

\subsection{Forward Model: from an IMF and Growth Law to the LFs}
\label{subsec:forward_model}

Given a choice of the five model parameters
$(m_{\rm k}, s, s_{\rm f}, \alpha, \tau_0)$, we compute a deterministic forward model for the pair of luminosity functions
$\{\phi_{\rm OB}(G_{\rm abs}),\phi_{\rm emb}(L)\}$, where $\phi_{\rm emb}(L)$ denotes the LF of the whole embedded ionizing phase sampled by the RMS catalogue, including both massive YSOs and compact H\,{\sc ii} regions.
The model assumes a constant birthrate and a broken-power-law field IMF,
\begin{equation}
\xi(m_{\rm f}) \propto
\begin{cases}
m_{\rm f}^{-s}, & m_{\rm f}<m_{\rm k},\\
m_{\rm k}^{\,s_{\rm f}-s}\,m_{\rm f}^{-s_{\rm f}}, & m_{\rm f}\ge m_{\rm k},
\end{cases}
\end{equation}
together with the growth law
\begin{equation}
t_{\rm form}(m_{\rm f})=\tau_0\left(\frac{m_{\rm f}}{m_0}\right)^{\alpha},
\end{equation}
with $m_0=60\,M_\odot$.

For the OB-star LF we use a precomputed kernel bank based on MIST evolutionary tracks.
For each final stellar mass $m_{\rm f}$, the kernel stores the cumulative time spent in each $G_{\rm abs}$ bin along the corresponding track.
This allows the LF to be evaluated by direct integration over final mass.
In addition, the model accounts for the fact that a star may have already burned part of its main-sequence fuel during the formation phase.
For a given $(\alpha,\tau_0)$, this is implemented as a growth-dependent shift of the starting point along each evolutionary track, so that the kernel entering the OB-star LF depends on both the IMF parameters and the growth-law parameters.
The model OB-star LF is then obtained by integrating the IMF-weighted kernel over $m_{\rm f}$.

For the embedded-source LF we again use a deterministic construction.
For each final mass $m_{\rm f}$, the contribution of the source to the LF is weighted by the duration of the embedded phase,
\begin{equation}
t_{\rm emb}(m_{\rm f}) = f_{\rm emb}(m_{\rm f})\,t_{\rm form}(m_{\rm f}),
\end{equation}
where in the present implementation the embedded phase is identified with the whole growth interval, so $f_{\rm emb}=1$.
Assuming the linear growth law of equation~(\ref{eq:growth_law}), the instantaneous mass is uniformly distributed along the interval from the onset of the embedded ionizing phase to $m_{\rm f}$.
Using the MIST-based mass--luminosity relation, we precompute for each $m_{\rm f}$ the fraction of the growth time spent in each luminosity bin.
The embedded-source LF is then obtained by integrating these contributions over the IMF.

Both model LFs are therefore deterministic functions of
$(m_{\rm k}, s, s_{\rm f}, \alpha, \tau_0)$.
The only additional free parameter is an overall normalization $A$, proportional to the birthrate amplitude.
Rather than scanning over $A$, we solve for its best-fitting value analytically at each point of the 5D parameter grid.
If $m_i$ denotes the concatenated model LF vector for unit normalization, $d_i$ the observed LF vector, and $\sigma_i$ the corresponding observational uncertainties, then
\begin{equation}
A_\star =
\frac{\sum_i m_i d_i/\sigma_i^2}
     {\sum_i m_i^2/\sigma_i^2},
\label{eq:Astar}
\end{equation}
and the corresponding minimum
\begin{equation}
\chi^2_{\rm min}=
\sum_i \frac{(A_\star m_i-d_i)^2}{\sigma_i^2}.
\label{eq:chi2min_joint}
\end{equation}
Thus the scanned parameter space is five-dimensional, not six-dimensional.

\subsection{Inference Pipeline}
\label{subsec:inference_pipeline}

The inference is performed with a joint 5D grid scan over the full parameter set
$(m_{\rm k}, s, s_{\rm f}, \alpha, \tau_0)$.
At each grid point, we compute the OB-star LF and the embedded-source LF as described above, concatenate the model predictions over the selected bins of the two observed LFs, solve analytically for the best-fitting normalization $A_\star$ from equation~(\ref{eq:Astar}), and evaluate the joint statistic from equation~(\ref{eq:chi2min_joint}).

The result of the scan is the 5D array
\begin{equation}
\chi^2(m_{\rm k}, s, s_{\rm f}, \alpha, \tau_0),
\end{equation}
together with the corresponding array of best-fitting normalizations
$A_\star(m_{\rm k}, s, s_{\rm f}, \alpha, \tau_0)$.
From this grid we derive the global best-fitting model, one-dimensional profiled constraints, and the two-dimensional profiled confidence regions shown in the figures.
In particular, the constraints on the growth-law parameters shown in Figure~4 are obtained by profiling the full 5D $\chi^2$ grid over the three IMF parameters.

Throughout the scan we compare the model to the same observational LFs derived in Section~\ref{sec:LFs} and Appendix~\ref{app:lf_corrections}, including both the Galactic-structure correction and the empirical completeness correction.
The fit therefore uses simultaneously the shape and relative normalization of the OB-star and embedded-source LFs, which is what allows the two datasets together to constrain both the IMF shape and the growth law.

\bibliography{refs}{}
\bibliographystyle{aasjournalv7}


\end{document}